\documentclass[preprint,floats,aps,epsfig,nofootinbib,amssymb]{revtex4-1}
\usepackage{mathrsfs}

\usepackage{slashed}
\usepackage{graphicx,color}
\usepackage{dcolumn}
\usepackage{bm}
\usepackage{subfig}
\usepackage{graphicx}
\usepackage{amssymb}
\usepackage{stackrel}

\begin{document}

\title{Hybrid Dark Matter}

\author{Wei Chao}
\email{chaowei@bnu.edu.cn}
\affiliation{Center for Advanced Quantum Studies, Department of Physics, Beijing Normal University, Beijing, 100875, China}

\vspace{3cm}

\begin{abstract}

Dark matter can be produced in the early universe  via the freeze-in or freeze-out mechanisms. Both scenarios were  investigated in references,  but the production  of dark matters via the combination of these two mechanisms are not addressed. In this paper we propose a hybrid dark matter model where dark matters have two components with one component produced thermally and the other one produced non-thermally.  We present for the first  time the analytical calculation for the relic abundance of the Higgs portal hybrid dark matter, then we investigate constraints on the parameter space of the model from dark matter direct and indirect detection experiments.

\end{abstract}

\maketitle
\section{Introduction}

Various astrophysical observations has confirmed the existence of the dark matter, with which about 26.8\%~\cite{Ade:2015xua} of our universe is constituted.  
We know nothing about the dark matter except its gravitational effects. Typically the thermal history of the dark matter remains a mystery, which catalyzes various conjecture of the dark matter models, such as weakly interacting massive particles (WIMPs)~\cite{Jungman:1995df}, feebly interacting massive particles (FIMPs)~\cite{Hall:2009bx}, strongly interacting massive particles (SIMPs)~\cite{Hochberg:2014dra}, etc.
WIMPs, which carry electroweak scale mass and couple to the standard model (SM) with a strength approximate to that of the weak interaction, are produced thermally via the so-called freeze-out mechanism. 
On the contrary, the FIMPs, which interact very weakly with the SM and have never attained thermal equilibrium, are produced non-thermally by the so-called freeze-in mechanism.  
As a result, WIMPs are accessible by the underground direct detection experiments, while FIMPs are not. Considering that no dark matter signal has been observed in any direct detection experiments, one can not justify which mechanism is better. 

In this paper we investigate a scenario where the dark matter has two components with one component produced by the freeze-out mechanism and the other one produced by the freeze-in mechanism. 
We dub this new dark matter scenario as the hybrid dark matter. 
As an illustration, we study the Higgs portal scalar hybrid dark matter model in detail, where the WIMP $\phi$ couples to the SM Higgs and the FIMP $\varphi$ couples to the $\phi$ with the quartic interaction.  
The relevant Lagrangian takes the form: 
\begin{eqnarray}
{\cal L} = -\mu^2 H^\dagger H + \lambda ( H^\dagger H)^2 +{1\over 2 } m_\phi^2 \phi^2 + {1\over 2 } m_\varphi^2 \varphi^2 + {1\over 2 }\lambda_\phi^{}  \phi^2 (H^\dagger H ) + {1\over 4 } \lambda_\varphi^{} \varphi^2 \phi^2  \; ,
\end{eqnarray}
where $\lambda_\varphi^{} \ll \lambda_\phi^{}$. Notice that there might be $\varphi^2 H^\dagger H$ term in the Lagrangian, here we have assumed that its coupling is negligibly small($<\lambda_\varphi$) for simplification. 
We show that the relic density of the FIMP can be derived analytically with a perturbative method. 
Numerical result shows that the relic density of the WIMP is not affected by that of the FIMP, even though it is produced from the WIMP.  
We further study constraints on the model from dark matter direct detection and indirect detection experiments.  
It shows that the Higgs portal hybrid dark matter model is excluded by direct detections for $m_\phi<479~{\rm GeV}$ except for $m_\phi \sim m_h/2$ (the resonant regime) and $m_\phi\sim m_h$ where the annihilation channel $\phi \phi \to hh$ is activated.  
For the alive dark matter mass range, the coupling strength $\lambda_\phi$ is further constrained by indirect detections. We show that  the Fermi Large Area Telescope (Fermi-LAT)~\cite{Ackermann:2015zua,Ackermann:2013uma} and  the High Energy Stereoscopic System (H.E.S.S.)~\cite{Abdallah:2016ygi} constrain  $\lambda_\phi$ to ${\cal O}(1)$.  The coupling strength of the FIMP is not constrained by direct and indirect detection experiments in this model. 

The remains of the paper is organized as follows:  In sections II we calculate the Boltzmann equations analytically and give numerical illustrations for the dark matter relic density. In section III we focus on constrains from both the direct  and the indirect detection experiments. The last part is concluding remarks, 

\section{Relic density}

We evaluate the relic density of the hybrid dark matter in this section. 
The evolution of the phase space distribution is controlled by the Boltzmann equations~\cite{Griest:1990kh}. 
After some manipulation, the Boltzmann equations can be written in the term of number densities, which take the following form: 
\begin{eqnarray}
\dot{n}_\phi +3Hn_\phi&=& -\langle \sigma v \rangle_\phi (n_\phi^2 -n_{\phi,{\rm eq}}^2) -\langle \sigma v\rangle_\varphi n_\phi^2 \label{eq1} \\
\dot{n}_\varphi+3Hn_\varphi &=&\langle \sigma v\rangle_\varphi n_\phi^2  \label{eq2}
\end{eqnarray}
where $n_\phi$ and $n_\varphi$ are number densities of $\phi$ and $\varphi$ respectively and the thermal average of the reduced annihilation cross section $\langle \sigma v \rangle_\varphi$ can be written as 
\begin{eqnarray}
\langle \sigma v\rangle_\varphi  ={1 \over 2^{11} \pi^7 T m_\phi^4 K_2^2 (m_\phi/T)} \int_{{\rm max} (4m_\phi^2, 4m_\varphi^2)} P_{\phi \phi}^{} P_{\varphi \varphi}^{} |{\cal M}|^2  {1\over \sqrt{s}} K_1^{}   \left(  {\sqrt{s} \over T}\right)  ds  
\end{eqnarray}
with $|{\cal M}|^2 =\lambda^2 $, and 
\begin{eqnarray}
P_{ij}= {1\over 2 \sqrt{s}} \sqrt{[s-(m_i+m_j)^2 ] [s-(m_i-m_j)^2]}
\end{eqnarray}
which is the momentum of the ``i" and ``j" particles in the frame of the center of the mass. 
$\langle  \sigma v\rangle_\phi$ is the same as the that of the conventional Higgs portal scalar dark matter model. We list in the appendix various annihilation cross sections of $\phi$.
Eqs.~(\ref{eq1}) and (\ref{eq2}) can be simplified in terms of the yield $Y\equiv n/s$, using the relation ${\dot{T}=-HT}$, 
where $H(\equiv1.66\sqrt{g_*} T^2 /M_{pl})$ is the Hubble constant with $g_*$ the effective degrees of freedom and with $M_{pl} (\equiv 1.22\times 10^{19}~\text{GeV})$ the Planck mass; $s(\equiv 2\pi^2 g_{*}^sT^3/45)$ is the entropy density and  $g_*^s $ can be replaced by $g_*$ for the most history of the Universe.

The total relic density of the hybrid dark matter can be written as
\begin{eqnarray}
\Omega h^2 = 2.82\cdot 10^8  \times \left(m_\phi^{}  Y_\phi^{} + m_\varphi^{} Y_\varphi^{}  \right )\; ,
\end{eqnarray}
where $Y_\phi^{}$ and $Y_\varphi^{}$ are derived from Eq. (\ref{eq1}) and Eq.~(\ref{eq2}). 

$Y_\varphi$ can be solved analytically as will be discussed in the following:
\begin{itemize}

\item For  $m_\phi < m_\varphi $,  the process $\phi +\phi \to \varphi +\varphi$ will be kinematically forbidden before the freeze-out of $\phi$.  
In this case,  the Eq.(\ref{eq2}) can be simplified  as 
\begin{eqnarray}
{d Y^0_\varphi \over d T} \approx  {1\over s H T} {\lambda^2 m_\varphi^2 T^2 \over 128 \pi^5} K_1^2 \left({m_\varphi \over T}\right) \; , 
\end{eqnarray}
where $Y_\varphi^0$ is the leading term of $Y_\varphi^{}$ by neglecting the mass of $\phi$.
Integrating with respect to the temperature, one has
\begin{eqnarray}
Y_\varphi ^{(0)} \approx { 45 \sqrt{\pi } M_{pl} \mathbb{\lambda}^2\over 1.66 \times 1024 \pi^7 \sqrt{g_*^{}} g_*^s m_\varphi  } G^{3,1}_{2,4} \left( {m_\varphi^2\over T^2} \left |\begin{array}{cccc}  1, & 2~ &&\cr {1\over 2},& {3\over 2},& {5\over 2}, & 0   \end{array}\right.\right)
\end{eqnarray}
where the rightest function is  the Meijer G-function.\footnote{ A general definition of the Meijer G-function is given by~\cite{MeijerGfunction}  
\begin{eqnarray}  
G^{m,n}_{p,q} \left( z \left |\begin{array}{ccc} a_1,&\cdots,& a_p \cr b_1, &\cdots,&b_q  \end{array} \right. \right)={1\over 2 \pi i } \int {\Pi_{j=1}^m \Gamma(b_j-s) \Gamma_{j=1}^n (1-a_j+s) \over \Pi_{j=m+1}^q \Gamma(1-b_j+s) \Pi_{j=n+1}^p \Gamma(a_j-s)} z^s ds \; , \nonumber
\end{eqnarray} 
where $\Gamma$ denotes the gamma function.}

Similarly, one can derive  the  higher order corrections to the $Y_\varphi$:
\begin{eqnarray}
Y_\varphi^{(1)}  &=& -{ 45 \sqrt{\pi } M_{pl} \mathbb{\lambda}^2\over 1.66 \times 1024 \pi^7 \sqrt{g_*^{}} g_*^s m_\varphi  }    {m_\phi^2 \over2 m_\varphi^2} G^{3,1}_{2,4} \left( {m_\varphi^2\over T^2} \left |\begin{array}{cccc}  1, & 3~ &&\cr {3\over 2},& {3\over 2},& {5\over 2}, & 0   \end{array}\right.\right)  \\
Y_\varphi^{(2)}  &=&  -{ 45 \sqrt{\pi } M_{pl} \mathbb{\lambda}^2\over 1.66 \times 1024 \pi^7 \sqrt{g_*^{}} g_*^s m_\varphi  }    {m_\phi^4 \over 8 m_\varphi^4} G^{3,1}_{2,4} \left( {m_\varphi^2\over T^2} \left |\begin{array}{cccc}  1, & 4~ &&\cr {3\over 2},& {5\over 2},& {5\over 2}, & 0   \end{array}\right.\right)
\end{eqnarray}

\item For $m_\varphi<m_\phi$,  $Y_\varphi$ can be calculated segmentally.  Assuming $x_F^\phi$ being the freeze-out parameter of  $\phi$, $Y_\varphi$ can be calculated perturbatively using the method developed above for $x<x_F^\phi$.  
The final expression of $Y_\varphi$ can then be written as 
\begin{eqnarray}
Y_\varphi =Y_\phi^2 \int_{x_F^\phi}^\infty { \langle \sigma v \rangle s \over H x} dx  + Y_\varphi^{} \left|_{x=x_F^\phi}  \right. \approx { \sqrt{g_\star} M_{pl} \lambda^2 Y_\phi^2 \over 74.7 \times 2^{12} \pi^5 x_F^\phi m_\phi } + Y_\varphi^{} \left|_{x=x_F^\phi} \right. \;  .
\end{eqnarray}
Apparently the first term on the rightest side, that is too much small, can be neglected and $Y_\varphi \approx Y_\varphi |_{x=x_F^\phi}$.

\end{itemize}

\begin{figure*}[t]
\begin{center}
  \includegraphics[width=0.45\textwidth]{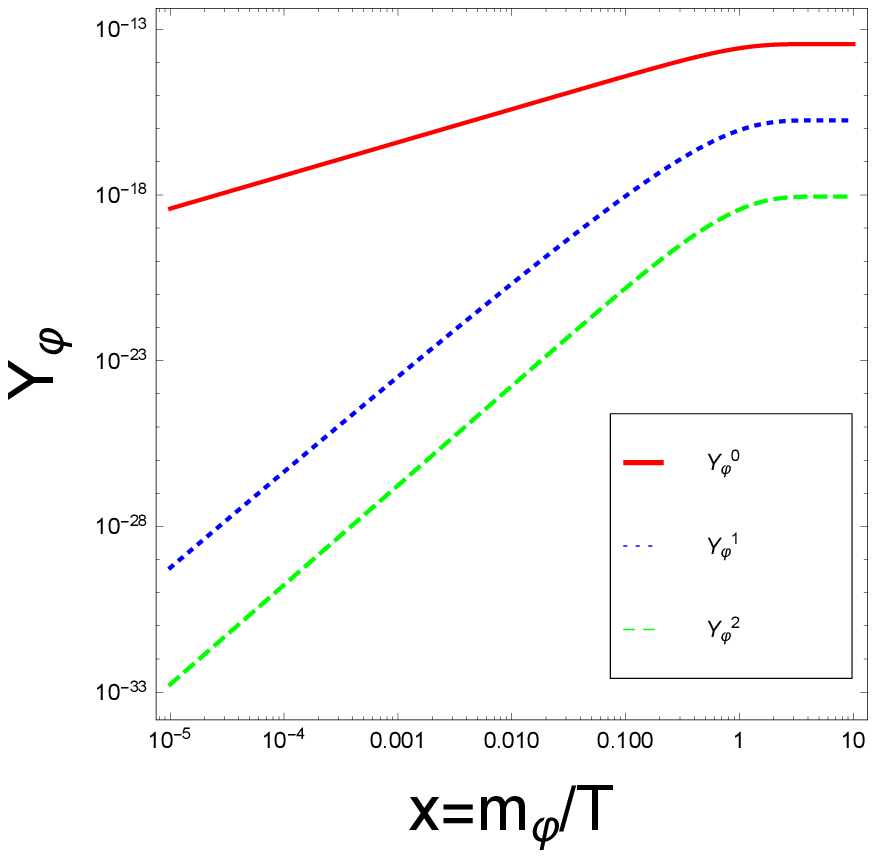} 
  \hspace{0.5cm}
    \includegraphics[width=0.45\textwidth]{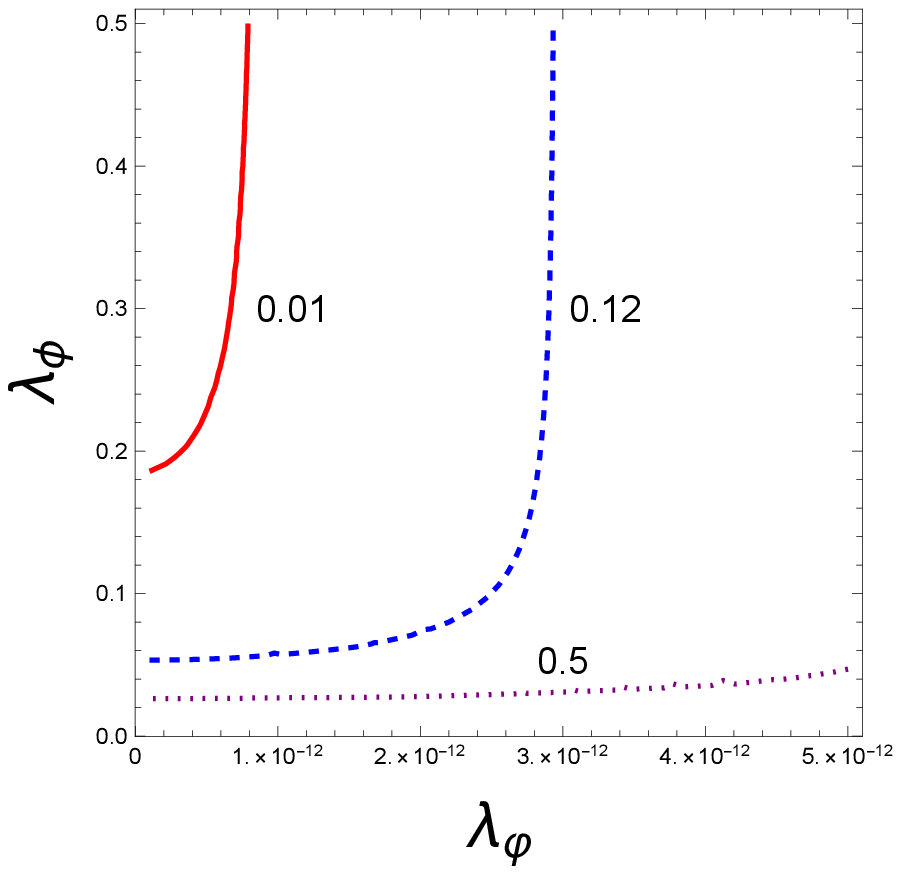} 
  \end{center}
\caption{Left panel: Illustrations of $Y_\varphi^0$, $Y_\varphi^1$ and $Y_\varphi^2$ as the function of  $x(\equiv m_\varphi/T)$ by setting $m_\varphi=1~\text{TeV}$, $m_\phi=200~\text{GeV}$ and $\lambda =10^{-11}$; right-panel: contours of the total dark matter relic density in the $\lambda_\varphi-\lambda_\phi$ plane by setting $m_\varphi=1~\text{TeV}$ and $m_\phi=200~\text{GeV}$.
\label{fig:y}
}

\end{figure*}

We continue to comment on the calculation $Y_\phi$, which is the same as the case of conventional Higgs portal~\cite{Patt:2006fw}. 
The thermal average of the reduced annihilation cross sections can be calculated analytically by expanding $s$ as $4m^2 + m^2 v^2 + 3/4 m^2 v^4$ in the laboratory frame. 
We approximate $\langle \sigma v \rangle$ with the non-relativistic expansion $\langle \sigma v \rangle = a + b \langle v^2 \rangle $ where  $v=v_{\text{lab}}$, then solve the Boltzmann equation analytically. 
%
%

For numerical illustrations, we show in the left-panel of the Fig.~\ref{fig:y} the $Y_\varphi^{(i)}$ as the function of $m_\varphi/T$ by setting $m_\varphi=1~\text{TeV}$, $m_\phi=200~\text{GeV}$ and $\lambda =10^{-11}$.
The solid, dotted and dashed lines correspond to $Y_\varphi^{(0)}$, $Y_\varphi^{(1)}$ and $Y_\varphi^{(2)}$ respectively.
As one can see $Y_\varphi^{(i)}$($i=0,~1$)  is about three orders larger than $Y_\varphi^{i+1} $ for ${\cal O}(x)\sim1$ and  $Y_\varphi^{(0)}$ is a good approximation.
We show in the right-panel of the Fig.~\ref{fig:y} contours of the total dark matter relic density in the $\lambda_\varphi-\lambda_\phi$ plane by setting $m_\varphi=1~\text{TeV}$ and $m_\phi=200~\text{GeV}$.  The solid, dashed and dotted lines correspond to $\Omega h^2 =0.01,0.12$ and $0.5$, respectively.  

\section{Direct and indirect detections}

The  hybrid dark matter contains two components: the FIMP and the WIMP. 
A FIMP could be detected at the LHC~\cite{Hessler:2016kwm}, but $\varphi$  in our model can not,  as it only couple to the WIMP, so that we mainly discuss signatures of $\phi$ in this section. 
It can be detected in direct detection experiments, which detect the energy imparted into nuclei in underground laboratories by collisions with the WIMP, and indirect detection experiments, which detect the fluxes of cosmic rays from the annihilation of the  WIMP that are gravitationally bound to the Galactic halo.  
For collider signatures of the Higgs portal dark matter, we refer the reader to Refs.~\cite{Djouadi:2012zc,Djouadi:2011aa,No:2013wsa,Chao:2016vfq} for detail, and we will not revisit them in this paper.

\begin{figure*}[t]
\begin{center}
  \includegraphics[width=0.47\textwidth]{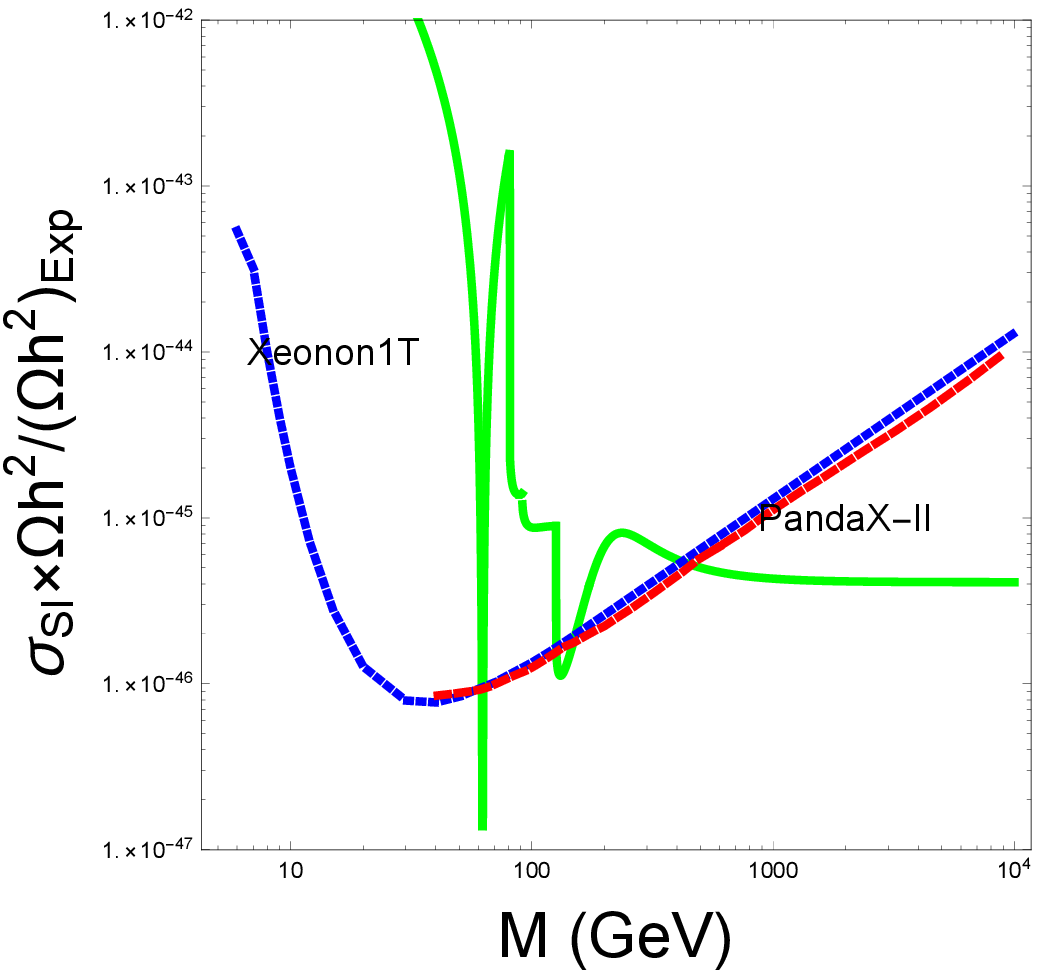} 
   \includegraphics[width=0.445\textwidth]{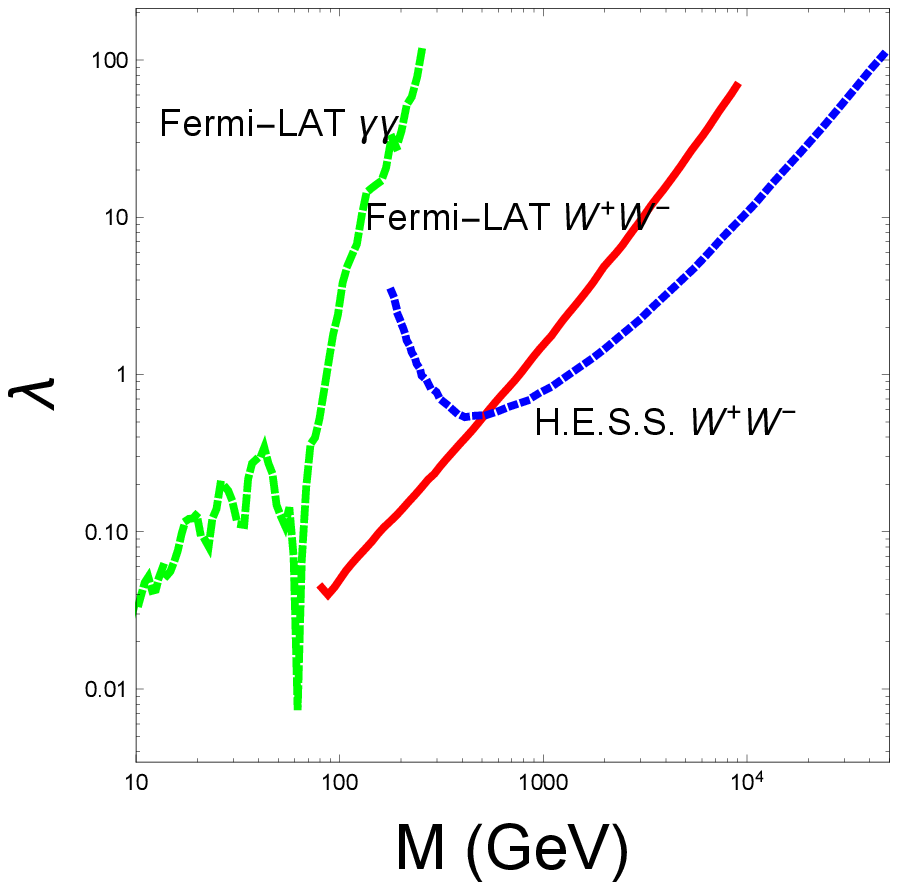} 
  \end{center}
\caption{Left-panel: The rescaled direct detection cross section as the function of the dark matter mass, the dotted  and dashed lines are constraints of Xenon1T and PandaX-II respectively; Right-panel:  Constraints from the dark matter indirect detections in the $M_\phi$ - $\lambda_\phi$ plane, the solid,  dashed and dotted lines are constraints from Fermi-LAT  $W^+W^-$ result, Fermi-LAT gamma-ray spectral lines result and H.E.S.S. $W^+W^-$ result, respectively.
\label{fig:DD}
}

\end{figure*}

The spin-independent direct detection cross section can be written as
\begin{eqnarray}
\sigma_{\rm SI}^{} = {\lambda_\phi^2  f_n^2 \over 4 \pi } { \mu^2 m_n^2 \over m_h^4 m_\phi^2 } \; ,
\end{eqnarray} 
where $\langle n | \sum m_q \bar q q | n\rangle =f_n^{}  m_n^{}$ and $f_n\approx 0.287$~\cite{Ellis:2000ds}, $\mu$ is the reduced mass of the DM-nucleon system: $\mu = m_n m_\phi /(m_n+ m_\phi)$. 
$\sigma_{\rm SI}$ needs to be multiplied by the factor, $\Omega h^2 |_\phi^{}  / \Omega h^2|_{\rm Exp } $, when compared with the experimental limit. 
It should be mentioned that the rescaled direct detection cross section is independent of the coupling $\lambda_\phi$. 

We show in the Fig.~\ref{fig:DD} the rescaled direct detection cross section as the function of the dark matter mass where the dotted and dashed lines are constraints of Xenon1T~\cite{Aprile:2017iyp} and PandaX-II~\cite{Cui:2017nnn}, respectively. 
Apparently the hybrid dark matter model is excluded by  direct detections for  $m_\phi<479~{\rm GeV}$ except for $m_\phi \sim m_h/2$ (the resonant region) and $m_\phi\sim m_h$ where the annihilation channel $\phi \phi \to hh$ is kinematically permitted.

The indirect detection of the WIMP tries to observe the cosmic rays coming from the annihilation of the WIMP pairs whenever the reduced annihilation cross section is not s-wave suppressed. For the hybrid dark matter considered in this paper, $\phi$ may have signals in the $\gamma \gamma$ and $W^+W^-$ channels. The differential gamma-ray flux from the annihilation of $\phi\phi$ is:
\begin{eqnarray}
{d \Phi_X \over d E d \Omega} = {1\over 8\pi} { \langle \sigma v\rangle_{} \over  m_\phi^2 } {d N_X \over d E} { d J_{\text{ann} } \over d \Omega}
\end{eqnarray}
where ${dN_X /d E}$~\cite{Cirelli:2010xx} is the differential `X'-ray yield per annihilation and $d J_{\text{ann}}/d\Omega$~\cite{Navarro:1995iw} is the integration of the square of the dark matter mass density along the line-of-sight. 
We will assume the dark matter mass density profile to be the Navarro-Frenk-While profile~\cite{Navarro:2008kc}.
The non-observation of any excess in gamma rays or $W^+W^-$ puts an upper limit on the reduced annihilation cross section $\langle \sigma v \rangle_{\gamma \gamma/W^+W^-}^{}$, which can be written as
\begin{eqnarray}
\langle \sigma v \rangle_{\gamma\gamma} &=& {\lambda_\phi^2 v^2 m_h   \Gamma_{h\to \gamma \gamma}\over 2m_\phi^2 [(4m_\phi^2 -m_h^2) + \Gamma_h^2 m_h^2 ]}   \\
\langle \sigma v\rangle_{W^+W^-} &= & {\lambda_\phi^2 m_W^4 \sqrt{m_\phi^2 -m_W^2 } \over 8 \pi m_\phi^3 [(4m_\phi^2 -m_h^2)^2 +m_h^2 \Gamma_h^2]} \left( 4{ m_\phi^4 \over m_W^4 } - 4 {m_\phi^2 \over m_W^2 } +3\right)
\end{eqnarray}
where $v$, $m_h$ and $\Gamma_h$ are respectively the vacuum expectation value, the mass and the total decay rate of the SM Higgs. 

We show in the right-panel of the Fig.~\ref{fig:DD} constraints of indirect detections in the $m_\phi^{}-\lambda_\phi^{}$ plane.  
The solid line is given by the Fermi-LAT~\cite{Ackermann:2015zua}, which searched for dark matter annihilation from the milky way dwarf spheroidal galaxies. 
The dotted line is given by the H.E.S.S.~\cite{Abdallah:2016ygi}, which searched for dark matter annihilation from the inner 300 pc of the Milky Way.   
Both the solid and dotted lines arise from the upper limits on the dark matter annihilation cross sections in the $W^+W^-$ channel. 
The dashed line is given by the Fermi-LAT~\cite{Ackermann:2013uma}, which searched for $\gamma$-ray spectral lines in 5$\sim$300 GeV. 
The results of $W^+W^-$ search shows that the constraint of  the Fermi-LAT is stronger for low dark matter mass, while that of the H.E.S.S. is stronger for high mass dark matter, and $\lambda_\phi$ should be blow ${\cal O}(1)$ for  $\phi$ at the TeV scale.

\section{Conclusion}

The nature of the dark matter is  a long standing problem to be addressed.
It makes sense to investigate various possibilities of the dark matter especially when it is going to be discovered in laboratories.  
In this paper we proposed the hybrid dark matter model where the dark matters are composed by a WIMP and a FIMP.
Taking the scalar Higgs portal hybrid dark matter as an example, we calculated the relic density of the FIMP analytically for the first time, and studied the signature of the WIMP in direct and indirect detection experiments.  
This work set a new precedent on the investigation of the multi-component dark matters, that come from different thermal histories.

\begin{acknowledgments}
This work was supported by the National Natural Science Foundation of China under grant No. 11775025 and the Fundamental Research Funds for the Central Universities under grant No. 2017NT17.
\end{acknowledgments}

\appendix

\section{annihilation cross sections }

\begin{eqnarray}
\sigma (\phi \phi \to WW/ZZ)   &=&  {\lambda_1^2  m_V^4 \over  4 \pi s  [(s-m_h^2 )^2 + m_h^2 \Gamma_h^2 ] }  \sqrt{ s-4m_V^2 \over s-4 m^2} \left( { {s^2 \over 4 m_V^4 } -{ s \over m_V^2 } +3 } \right) \\
\sigma (\phi \phi \to \gamma \gamma /Z \gamma)   &=& {\lambda_1^2 v^2 m_h^{}  \over \sqrt{ s (s-4m^2 )} }  {1 \over (s-m_h^2 )^2 + m_h^2 \Gamma_h^2 } \Gamma_{h \to \gamma \gamma /Z \gamma}^{}  \\
\sigma(\phi \phi \to f \bar f ) &=& { \lambda_1^2 m_f^2 (s-4m_f^2)^{3/2} \over 8 \pi s \sqrt{s-4m^2 }}  {1 \over (s-m_h^2)^2 + m_h^2 \Gamma_h^2}
\end{eqnarray}

\begin{eqnarray}
\sigma(\phi \phi \to hh ) =
{ \sqrt{s-4m_h^2 } \over 8\pi s \sqrt{s-4m^2} }\left\{ \frac{2 \lambda _1^4 v^4}{m^2 \left(s-4 m_h^2\right)+m_h^4}+\left(\frac{6 \lambda  \lambda _1 v^2}{s-m_h^2}+\lambda _1\right)^2 + \right.  \nonumber \\
\left.\frac{8 \lambda _1^3 v^2  \left [m_h^2 \left(\lambda _1 v^2-3 s-12  \lambda  v^2 \right)+2 m_h^4+s \left(s+6 \lambda  v^2-\lambda _1 v^2\right)\right] }{\sqrt{\left(s-4 m^2\right) \left(s-4 m_h^2\right)} \left(2 m_h^4+s^2 -3 s m_h^2\right) \tanh \left[ { \sqrt{\left(s-4 m^2\right) \left(s-4 m_h^2\right)} \over \left(2 m_h^2-s\right)}\right] } \right\}
\end{eqnarray}


\end{document}